\documentclass[pdftex,twocolumn,epjc3]{svjour3}          % twocolumn

\smartqed  % flush right qed marks, e.g. at end of proof

\RequirePackage{mathptmx}      % use Times fonts if available on your TeX system
\usepackage{flushend}
\usepackage[numbers,sort&compress]{natbib}
\usepackage[colorlinks,citecolor=blue,urlcolor=blue,linkcolor=blue]{hyperref}
\usepackage{graphicx}% Include figure files
\usepackage{subfig}
\usepackage{dcolumn}% Align table columns on decimal point
\usepackage{bm}% bold math
\usepackage{float}
\usepackage{xcolor}
\usepackage{amssymb}
\usepackage{amsmath}
\journalname{Eur. Phys. J. C}
\DeclareMathAlphabet{\mathcal}{OMS}{cmsy}{m}{n}
\DeclareSymbolFont{largesymbols}{OMX}{cmex}{m}{n}
\begin{document}

\title{Testing backreaction effects with observational Hubble parameter data}

\author{Shu-Lei Cao\thanksref{addr1}
        \and
        Huan-Yu Teng\thanksref{addr1} %etc.
        \and
        Hao-Yi Wan\thanksref{addr1,addr3}
        \and
        Hao-Ran Yu\thanksref{addr2}
        \and
        Tong-Jie Zhang\thanksref{e1,addr1,addr4}
}

\thankstext{e1}{e-mail: tjzhang@bnu.edu.cn}

\institute{Department of Astronomy, Beijing Normal University, Beijing, 100875, China\label{addr1}
          \and
          Tsung-Dao Lee Institute, Shanghai Jiao Tong University, 800 Dongchuan Road, Shanghai, 200240, China\label{addr2}
          \and
          National Astronomical Observatories, Chinese Academy of Sciences, Beijing 100012, China\label{addr3}
          \and
          School of Information Management, School of Physics and Electric Information, Shandong Provincial Key Laboratory of Biophysics, Dezhou University, Dezhou 253023,China\label{addr4}
}

\date{Received: date / Accepted: date}

\maketitle

\begin{abstract}
The spatially averaged inhomogeneous Universe includes a kinematical backreaction term ${\mathcal{Q}}_{\mathcal{D}}$ that is relate to the averaged spatial Ricci scalar ${\langle \mathcal{R}} \rangle_{\mathcal{D}}$ in the framework of general relativity. Under the assumption that ${\mathcal{Q}}_{\mathcal{D}}$ and ${\langle \mathcal{R}} \rangle_{\mathcal{D}}$ obey the scaling laws of the volume scale factor $a_{\mathcal{D}}$, a direct coupling between them with a scaling index $n$ is remarkable. In order to explore the generic properties of a backreaction model for explaining the accelerated expansion of the Universe, we exploit two metrics to describe the late time Universe. Since the standard FLRW metric cannot precisely describe the late time Universe on small scales, the template metric with an evolving curvature parameter $\kappa_{\mathcal{D}}(t)$ is employed. However, we doubt the validity of the prescription for $\kappa_{\mathcal{D}}$, which motivates us apply observational Hubble parameter data (OHD) to constrain parameters in dust cosmology. First, for FLRW metric, by getting best-fit constraints of $\Omega^{{\mathcal{D}}_0}_m = 0.25^{+0.03}_{-0.03}$, $n = 0.02^{+0.69}_{-0.66}$, and $H_{\mathcal{D}_0} = 70.54^{+4.24}_{-3.97}\ {\rm km \ s^{-1} \ Mpc^{-1}}$, the evolutions of parameters are explored. Second, in template metric context, by marginalizing over $H_{\mathcal{D}_0}$ as a prior of uniform distribution, we obtain the best-fit values of $n=-1.22^{+0.68}_{-0.41}$ and ${{\Omega}_{m}^{\mathcal{D}_{0}}}=0.12^{+0.04}_{-0.02}$. Moreover, we utilize three different Gaussian priors of $H_{\mathcal{D}_0}$, which result in different best-fits of $n$, but almost the same best-fit value of ${{\Omega}_{m}^{\mathcal{D}_{0}}}\sim0.12$. Also, the absolute constraints without marginalization of parameter are obtained: $n=-1.1^{+0.58}_{-0.50}$ and ${{\Omega}_{m}^{\mathcal{D}_{0}}}=0.13\pm0.03$. With these constraints, the evolutions of the effective deceleration parameter $q^{\mathcal{D}}$ indicate that the backreaction can account for the accelerated expansion of the Universe without involving extra dark energy component in the scaling solution context. Nevertheless, the results also verify that the prescription of $\kappa_{\mathcal{D}}$ is insufficient and should be improved.
\end{abstract}

\section{Introduction}

The universe is homogeneous and isotropic on very large scales. According to Einstein's general relativity, one can obtain a homogeneous and isotropic solution of Einstein's field equations, which is called Friedmann-Lemaitre-Robertson-Walker (FLRW) metric. Since on smaller scales, the universe appears to be strongly inhomogeneous and anisotropic, Larena et al. \cite{Larena2009} doubt that the FLRW cosmology describes the averaged inhomogeneous universe at all times. They assume that FLRW metric may not hold at late times especially when there are large matter inhomogeneities existed, even though it may be suitable at early times. Therefore they introduce a template metric that is compatible with homogeneity and isotropy on large scales of FLRW cosmology, and also contains structure on small scales. In other words, this metric is built upon weak, instead of strong, cosmological principle.

The observations of type \uppercase\expandafter{\romannumeral1}a supernovae (SNe \uppercase\expandafter{\romannumeral1}a) \cite{Perlmutter1999,Riess1998} suggest that the universe is in a state of accelerated expansion, which implies that there exists a latent component so called Dark Energy (DE) with negative pressure that causes the accelerated expansion of the universe. There are many scenarios proposed to account for the observations. The simplest one is the positive cosmological constant in Einstein's equations, which in common assumption is equivalent to the quantum vacuum. Since the measured cosmological constant is much smaller than the particle physics predicted, some other scenarios are proposed, such as the phenomenological models which explained DE as a late time slow rolling scalar field \cite{Copeland2006} or the Chaplygin gas \cite{Gorini2009}, and the modified gravity models. Recently, a new scenario \cite{Rasanen2004,Kolb2006} is raised to consider DE as a backreaction effect of inhomogeneities on the average expansion of the universe. Here we specifically focus on the backreaction model without involving perturbation theory.

According to Buchert \cite{Buchert2000}, the averaged equations of the averaged spatial Ricci scalar ${\langle \mathcal{R}} \rangle_{\mathcal{D}}$ and the `backreaction' term ${\mathcal{Q}}_{\mathcal{D}}$ can be solved to obtain the exact scaling solutions in which a direct coupling between ${\langle \mathcal{R}} \rangle_{\mathcal{D}}$ and ${\mathcal{Q}}_{\mathcal{D}}$ with a scaling index $n$ is significant. With that solution, a domain-dependent Hubble function (effective volume Hubble parameter) $H_{\mathcal{D}}$ can be expressed with the scaling index $n$ and the present effective matter density parameter ${{\Omega}_{m}^{\mathcal{D}_{0}}}$. Also, as mentioned in \cite{Larena2009}, the pure scaling ansatz is not what we expected in a realistic evolution of backreaction.

So as to explore the generic properties of a backreaction model for explaining the observations of the Universe, we, in this paper, exploit two metrics to describe the late time Universe. Although the template metric proposed by Larena et. al. \cite{Larena2009} is reasonable, the prescription of the so-called ``geometrical instantaneous spatially-constant curvature" $\kappa_{\mathcal{D}}$ is skeptical, based on the discrepancies between our results and theirs. Comparing the FLRW metric with the smoothed template metric, we use observational Hubble parameter data (OHD) to constrain the scaling index $n$ (corresponding to constant equation of state for morphon field $w^{\mathcal{D}}_{\Phi}$ \cite{Buchert2006}) and the present effective matter density parameter ${{\Omega}_{m}^{\mathcal{D}_{0}}}$ without involving perturbation theory. In the latter case, according to \cite{Ma2011}, we choose to marginalize over both the top-hat prior of $H_{{\mathcal{D}}_0}$ with a uniform distribution in the interval [50, 90] and three different Gaussian priors of $H_{{\mathcal{D}}_0}$, where we also obtain the absolute constraint results without the marginalization of the parameters. Combining both the FLRW geometry and the template metric with the backreaction model, we obtain the fine relation between effective Hubble parameter $H_{\mathcal{D}}$ and effective scale factor $a_{\mathcal{D}}$ by utilizing Runge-Kutta method to solve the differential equations of the latter, in order to acquire the link between $a_{\mathcal{D}}$ and effective redshift $z_{\mathcal{D}}$. At last, a conflict, as expected, arises. Our results show that it needs a higher instead of lower amount of backreaction to interpret the effective geometry, even though accelerated expansion of $a_{\mathcal{D}}$ still remains. The power law prescription of $\kappa_{\mathcal{D}}$ certainly need to be improved, since it only evolves from 0 to -1, which is insufficient. Of course, we should point out that the power law ansatz is not the realistic case and the results are expected to be inaccurate. For simplicity, we only deliberate the situation under the assumption of power-law ansatz here.

The paper is organized as follows. The backreaction context is demonstrated in Section \ref{sec:2}. In Section \ref{sec:3}, we introduce the template metric and computation of observables along with the effective Hubble parameter $H_{\mathcal{D}}$, and demonstrate how to relate effective redshift ${z_{\mathcal{D}}}$ to effective scale factor $a_{\mathcal{D}}$. We also refer to overall cosmic equation of state $w^{\mathcal{D}}_{\rm eff}$ \cite{Buchert2006} and how it differs from constant equation of state $w$. In Section \ref{sec:4}, according to the effective Hubble parameter, we apply OHD with both the FLRW metric and the template metric, and make use of Metropolis-Hastings algorithm of the Markov-Chain-Monte-Carlo (MCMC) method and mesh-grid method, respectively, to obtain the constraints of the parameters. In former case, we employ the best-fits to illustrate the evolutions of ${q}^{\mathcal{D}}$, $w^{\mathcal{D}}_{\rm eff}$, $\kappa_{\mathcal{D}}$ and density parameters. In latter case, we test the effective deceleration parameter with the best-fit values. After analysis of the results in Section \ref{sec:4}, we summarize our conclusion and discussion in Section \ref{sec:5}.

We use the natural units $c = 1$ throughout the paper, and assign that Greek indices such as $\alpha$, $\mu$ run through 0...3, while Latin indices such as $i, j$ run through 1...3.

\section{The bacreaction model}
\label{sec:2}

Buchert \cite{Buchert2000} introduced a model of dust cosmologies, which leads to two averaged equations, the averaged Raychaudhuri equation
\begin{equation}\label{eq:1}
  3\frac{\ddot{a}_{\mathcal{D}}}{a_{\mathcal{D}}}+4\pi G\langle\varrho\rangle_{\mathcal{D}}-\Lambda={\mathcal{Q}}_{\mathcal{D}} \  ,
\end{equation}
and the averaged Hamiltonian constraint
\begin{equation}\label{eq:2}
  3(\frac{\dot{a}_{\mathcal{D}}}{a_{\mathcal{D}}})^2-8\pi G\langle\varrho\rangle_{\mathcal{D}}+\frac{1}{2}\langle {\mathcal{R}}\rangle_{\mathcal{D}}-\Lambda=-\frac{{\mathcal{Q}}_{\mathcal{D}}}{2} \  ,
\end{equation}
where $\langle\varrho\rangle_{\mathcal{D}}$, $G$, and $\Lambda$ represent averaged matter density in the domain ${\mathcal{D}}$, gravitational constant, and cosmological constant, respectively. The over-dot represents partial derivative with respect to proper time $t$ here after. A effective scale factor is introduced via volume (normalized by the volume of the initial domain $V_{\mathcal{D}_i}$),
\begin{equation}\label{eq:3}
  a_{\mathcal{D}}(t)=(\frac{V_{\mathcal{D}}(t)}{V_{{\mathcal{D}}_i}})^{1/3} \  .
\end{equation}
The averaged spatial Ricci scalar $\langle {\mathcal{R}}\rangle_{\mathcal{D}}$ and the `backreaction' ${\mathcal{Q}}_{\mathcal{D}}$ are domain-dependent constants, which yet are time-dependent functions. The `backreaction' term is expressed as
\begin{equation}\label{eq:4}
  {\mathcal{Q}}_{\mathcal{D}}=2\langle \textbf{II} \rangle_{\mathcal{D}}-\frac{2}{3}\langle \textbf{I}\rangle^2_{\mathcal{D}}=
\frac{2}{3}\langle(\theta-\langle \theta\rangle_{\mathcal{D}})^2\rangle_{\mathcal{D}}-2\langle\sigma^2 \rangle_{\mathcal{D}}\  ,
\end{equation}
with two scalar invariants
\begin{equation}\label{eq:5}
  \textbf{I}=\Theta^l_{\ l}=\theta\  ,
\end{equation}
and
\begin{equation}\label{eq:6}
  \textbf{II}=\frac{1}{2}(\theta^2-\Theta^l_{\ k}\Theta^k_{\ l})=\frac{1}{3}\theta^2-\sigma^2\  ,
\end{equation}
where $\Theta_{ij}$ is the expansion tensor, with the trace-free symmetric shear tensor $\sigma_{ij}$, the rate of shear ${\sigma}^2=\frac{1}{2}{\sigma}^i_{\  j}{\sigma}^j_{\  i}$ and the expansion rate $\theta$.
Substituting Eq. (\ref{eq:1}) into Eq. (\ref{eq:2}), one can get

\begin{equation}\label{eq:7}
  (a^6_{\mathcal{D}}{\mathcal{Q}}_{\mathcal{D}})^\bullet+a^4_{\mathcal{D}}(a^2_{\mathcal{D}}\langle {\mathcal{R}}\rangle_{\mathcal{D}})^\bullet=0\  ,
\end{equation}
where the dot over the parentheses represents partial derivative over time $t$, and the scaling solutions are
\begin{equation}\label{eq:8}
  \langle {\mathcal{R}}\rangle_{\mathcal{D}}=\langle {\mathcal{R}}\rangle_{{\mathcal{D}}_i}a^n_{\mathcal{D}}\   ; {\mathcal{Q}}_{\mathcal{D}}={\mathcal{Q}}_{{\mathcal{D}}_i}a^m_{\mathcal{D}}\  ,
\end{equation}
where $n$ and $m$ are real numbers. As mentioned in \cite{Buchert2000}, there are two types of solutions to be considered.
The first type is $n=-2$ and $m=-6$, which corresponds to a quasi-FLRW universe at late times, i.e., backreaction is negligible. The second type is a direct coupling between $\langle {\mathcal{R}}\rangle_{\mathcal{D}}$ and ${\mathcal{Q}}_{\mathcal{D}}$, i.e., $m=n$, which reads
\begin{equation}\label{eq:9}
  \langle {\mathcal{R}}\rangle_{\mathcal{D}}=\langle {\mathcal{R}}\rangle_{{\mathcal{D}}_i}a^n_{\mathcal{D}}\  ,
\end{equation}
and
\begin{equation}\label{eq:10}
  {\mathcal{Q}}_{\mathcal{D}}=-\frac{n+2}{n+6}\langle {\mathcal{R}}\rangle_{{\mathcal{D}}_i}a^n_{\mathcal{D}}\  .
\end{equation}
A domain-dependent Hubble function is defined to be $H_{\mathcal{D}}=\dot {a_{\mathcal{D}}}/{a_{\mathcal{D}}}$, and dimensionless (`effective') averaged cosmological parameters are also given respectively by
\begin{equation}\label{eq:11}
  \Omega^{\mathcal{D}}_m=\frac{8\pi G\langle\varrho\rangle_{\mathcal{D}}}{3H^2_{\mathcal{D}}}\  ,
\end{equation}
\begin{equation}\label{eq:12}
  \Omega^{\mathcal{D}}_\Lambda=\frac{\Lambda}{3H^2_{\mathcal{D}}}\  ,
\end{equation}
\begin{equation}\label{eq:13}
  \Omega^{\mathcal{D}}_{\mathcal{R}}=-\frac{\langle {\mathcal{R}}\rangle_{\mathcal{D}}}{6H^2_{\mathcal{D}}}\  ,
\end{equation}
and
\begin{equation}\label{eq:14}
  \Omega^{\mathcal{D}}_{\mathcal{Q}}=-\frac{{\mathcal{Q}}_{\mathcal{D}}}{6H^2_{\mathcal{D}}}\  .
\end{equation}
Thus, according to Eq. (\ref{eq:2}), one can have
\begin{equation}\label{eq:15}
  \Omega^{\mathcal{D}}_m+\Omega^{\mathcal{D}}_\Lambda+\Omega^{\mathcal{D}}_{\mathcal{R}}+\Omega^{\mathcal{D}}_{\mathcal{Q}}=1\  .
\end{equation}
The components that are not included in Friedmann equation read
\begin{equation}\label{eq:16}
  \Omega^{\mathcal{D}}_X=\Omega^{\mathcal{D}}_{\mathcal{R}}+\Omega^{\mathcal{D}}_{\mathcal{Q}}\  .
\end{equation}
If $\Omega^{\mathcal{D}}_\Lambda=0$, then $\Omega^{\mathcal{D}}_X$ is considered to be the DE contribution, and usually dubbed as X-matter. Considering Eqs. (\ref{eq:9}) and (\ref{eq:10}), one can get
\begin{equation}\label{eq:17}
  \Omega^{\mathcal{D}}_X=-\frac{2\langle {\mathcal{R}}\rangle_{{\mathcal{D}}_i}a^n_{\mathcal{D}}}{3(n+6)H^2_{\mathcal{D}}}\  .
\end{equation}
Furthermore, one can easily obtain
\begin{equation}\label{eq:18}
  H^2_{\mathcal{D}}(a_{\mathcal{D}})=H^2_{{\mathcal{D}}_0}(\Omega^{{\mathcal{D}}_0}_m{a^{-3}_{\mathcal{D}}}+\Omega^{{\mathcal{D}}_0}_X{a^n_{\mathcal{D}}})\  ,
\end{equation}
where ${{\mathcal{D}}_0}$ denotes the domain at present time, and $a_{{\mathcal{D}}_0}=1$ here after.

In comparison with the deceleration parameter $q$ in standard cosmology, an effective volume deceleration parameter ${q}^{\mathcal{D}}$ is interpreted as
\begin{equation}\label{eq:19}
  q^{\mathcal{D}}= -\frac{\ddot{a}_{\mathcal{D}}}{a_{\mathcal{D}}}\frac{1}{H^2_{\mathcal{D}}}=\frac{1}{2}\Omega^{\mathcal{D}}_m+2\Omega^{\mathcal{D}}_{\mathcal{Q}}-\Omega^{\mathcal{D}}_\Lambda.
\end{equation}

\section{Effective geometry}
\label{sec:3}

\subsection{The template metric}

Larena et al. \cite{Larena2009} proposed the template metric (space-time metric) as follows,
\begin{equation}\label{eq:20}
  ^4g^{\mathcal{D}}=-{\rm d}t^2+L^2_{H_0}a^2_{\mathcal{D}}\gamma^{\mathcal{D}}_{ij}{\rm d}X^i\otimes {\rm d}X^j\  ,
\end{equation}
where $a_{{\mathcal{D}}_0}L_{H_0}=1/H_{{\mathcal{D}}_0}$ is introduced as the size of the horizon at present time, so that the coordinate distance is dimensionless, and the domain-dependent \emph{effective 3-metric} reads
\begin{equation}\label{eq:21}
  \gamma^{\mathcal{D}}_{ij}{\rm d}X^i\otimes {\rm d}X^j=(\frac{{\rm d}r^2}{1-\kappa_{\mathcal{D}}(t)r^2}+{\rm d}\Omega^2)\  ,
\end{equation}
with solid angle element ${\rm d}\Omega^2=r^2({\rm d}\theta^2+sin^2\theta {\rm d}\phi^2)$.
Under their assumption, the template 3-metric is identical to the spatial part of a FLRW space-time at any given time, except for the time-dependent scalar curvature. According to their discussion, $\kappa_{\mathcal{D}}$ must be related to $\langle {\mathcal{R}}\rangle_{\mathcal{D}}$, thus in analogy with a FLRW metric, the correlation can be given by
\begin{equation}\label{eq:22}
  \langle {\mathcal{R}}\rangle_{\mathcal{D}}=\frac{\kappa_{\mathcal{D}}(t)|\langle {\mathcal{R}}\rangle_{{\mathcal{D}}_0}|a^2_{{\mathcal{D}}_0}}{a^2_{\mathcal{D}}(t)}\  .
\end{equation}
Notice that this template metric does not need to be a dust solution of Einstein's equations, since Einstein's field equations are satisfied locally for any space-time metric. Nevertheless, this prescription is insufficient, as $\kappa_{\mathcal{D}}$ cannot be positive in this case, which is assertive and skeptical.
\subsection{Computation of Observables}

\begin{figure*}
\centering
\includegraphics[width=0.65\textwidth]{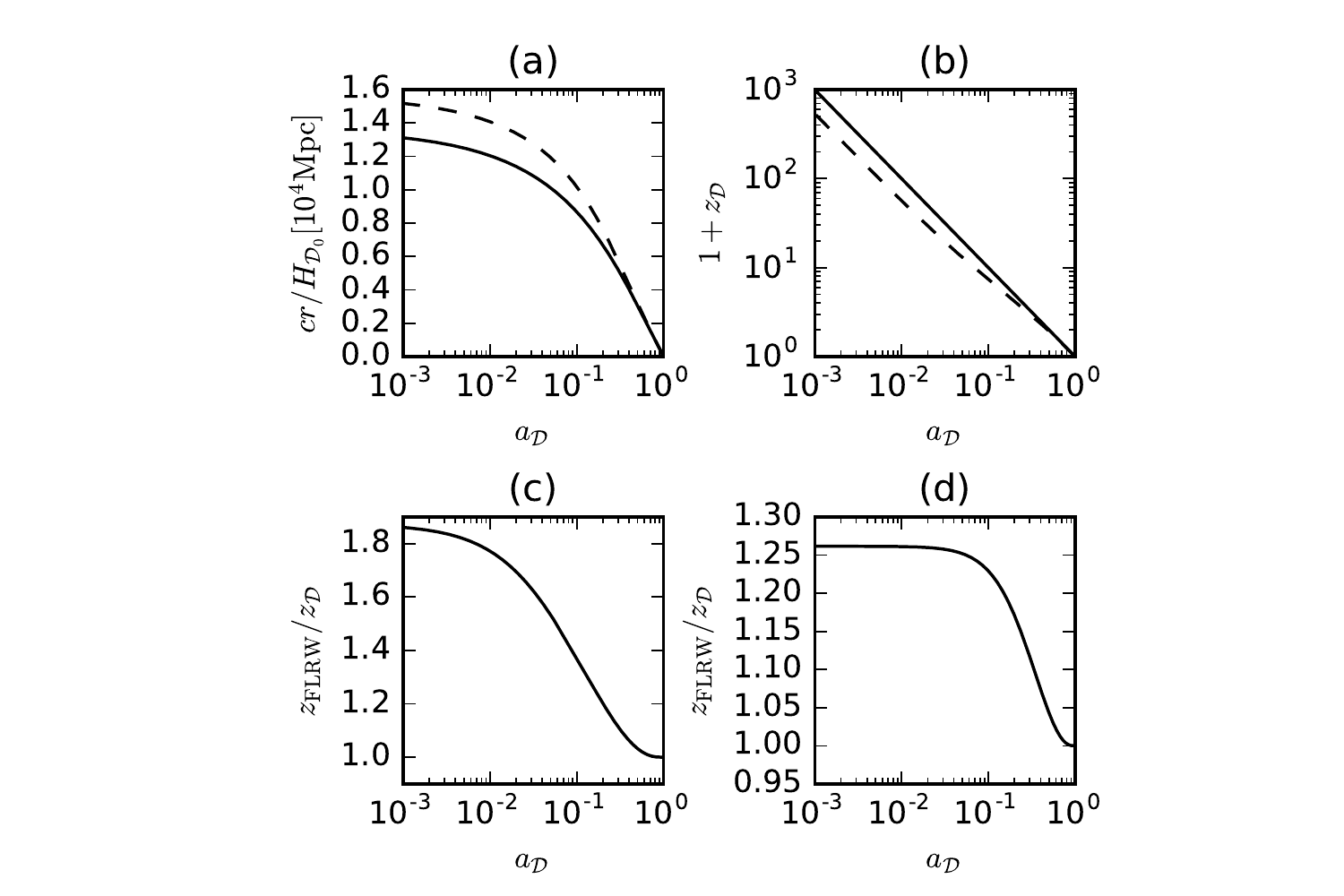}
\caption{The evolutions of the coordinate distance $cr/H_{{\mathcal{D}}_0}$ (panel (a)), the redshift (panel (b)) and the ratio between the FLRW redshift and the effective redshift, $1/(a_{\mathcal{D}}(1+z_{\mathcal{D}}))$, in the averaged model as function of the effective scale factor $z_{\mathcal{D}}$ with $n = -1$, $\Omega^{{\mathcal{D}}_0}_m = 0.3$, $H_{{\mathcal{D}}_0} = 70\ {\rm km \ s^{-1} \ Mpc^{-1}}$ (dashed line). The flat FLRW model with same parameters are the solid lines shown in panels (a) and (b), respectively. Panel (d) represents $1/(a_{\mathcal{D}}(1+z_{\mathcal{D}}))$ for Larena's best-fit averaged model with $n=0.12$, $\Omega^{{\mathcal{D}}_0}_m = 0.38$, $H_{{\mathcal{D}}_0} = 78.54\ {\rm km \ s^{-1} \ Mpc^{-1}}$.}\label{fig:1}
\end{figure*}

The computation of effective distances along the approximate smoothed light cone associated with the travel of light is very different from general that of distances \cite{Bonvin2006}. Firstly, an effective volume redshift $z_{\mathcal{D}}$ is defined as
\begin{equation}\label{eq:23}
  1+z_{\mathcal{D}}=\frac{(g_{\mu\nu}k^{\mu}u^{\nu})_S}{(g_{\mu\nu}k^{\mu}u^{\nu})_O}\  ,
\end{equation}
where the O and the S represent the evaluation of the quantities at the observer and source, respectively, $g_{\mu\nu}$ is the template effective metric, $u^{\mu}$ is the 4-velocity of the matter content ($u^{\mu}u_{\mu}=-1$) with respect to comoving reference, and $k^{\mu}$ is the wave vector of a light ray that travels from the source S to the observer O ($k^{\mu}k_{\mu}=1$). Normalizing the wave vector $(k^{\mu}u_{\mu})_O=-1$ and defining the scaled vector $\hat k^{\mu}={a^2_{\mathcal{D}}}k^{\mu}$, we can obtain the following relation
\begin{equation}\label{eq:24}
  1+z_{\mathcal{D}}=\left({a^{-1}_{\mathcal{D}}}\hat k^0\right)_S\  ,
\end{equation}
where $\hat k^0$ obeys the null geodesics equation $\hat k^{\mu}\nabla_{\mu}\hat k^{\nu}=0$, which can lead to
\begin{equation}\label{eq:25}
  \frac{1}{\hat k^0}\frac{{\rm d}{\hat k^0}}{{\rm d}a_{\mathcal{D}}}=-\frac{r^2(a_{\mathcal{D}})}{2(1-{\kappa_{\mathcal{D}}}(a_{\mathcal{D}})r^2(a_{\mathcal{D}}))}\frac{{{\rm d}{\kappa_{\mathcal{D}}}}(a_{\mathcal{D}})}{{\rm d}a_{\mathcal{D}}}\ .
\end{equation}
As light travels along null geodesic, we have
\begin{equation}\label{eq:26}
  \frac{{\rm d}r}{{\rm d}a_{\mathcal{D}}}=-\frac{H_{{\mathcal{D}}_0}}{a^2_{\mathcal{D}}H_{\mathcal{D}}(a_{\mathcal{D}})}\sqrt{1-\kappa_{\mathcal{D}}(a_{\mathcal{D}})r^2}\  ; r(1)=0\  ,
\end{equation}
which is slightly different from Eq. (30) of \cite{Larena2009}, since $r(0)=0$ they choose is actually $r(a_{{\mathcal{D}}_0}=1)=0$ (Note: this should be just a typo and does not affect the results.). By solving Eq. (\ref{eq:26}) we can get the coordinate distance $\bar r(a_{\mathcal{D}})$, and then substitutes it into Eq. (\ref{eq:25}) to find the relation between $z_{\mathcal{D}}$ and $a_{\mathcal{D}}$.
Combining equations in Section \ref{sec:3} and Section \ref{sec:2}, we can obtain
\begin{equation}\label{eq:27}
  \kappa_{\mathcal{D}}(a_{\mathcal{D}})=-\frac{(n+6)\Omega^{{\mathcal{D}}_0}_Xa^{n+2}_{\mathcal{D}}}{|(n+6)\Omega^{{\mathcal{D}}_0}_X|}\  ,
\end{equation}
\begin{equation}\label{eq:28}
  \frac{{\rm d}r}{{\rm d}a_{\mathcal{D}}}=-\sqrt{\frac{1-\kappa_{\mathcal{D}}(a_{\mathcal{D}})r^2}{\Omega^{{\mathcal{D}}_0}_m{a_{\mathcal{D}}}+\Omega^{{\mathcal{D}}_0}_X {a^{n+4}_{\mathcal{D}}}}} \quad ; \quad r(1)=0\  .
\end{equation}
Note that there are also typos in Eq. (41) of \cite{Larena2009}, and the correct one is Eq. (\ref{eq:28}). As shown in Fig. \ref{fig:1}, panels (a) and (d) are identical to Larena's Fig. 1, and panels (b) and (c) have same evolutionary trends but different limits on the left, where when $a_{\mathcal{D}}=10^{-3}$, $z_{\rm FLRW}/z_{\mathcal{D}}\approx1.86$ for our work here, but $z_{\rm FLRW}/z_{\mathcal{D}}\approx2.63$ for their work. Since panels (c) and (d) are obtained with the same method and just different equations, we are certain about the precisions of our results.

In DE context, a constant equation of state $w$ is correlated with the component $n$ as $w^{\mathcal{D}}=-(n+3)/3\  .$ However, as for the time-dependent curvature of the backreaction model, due to \cite{Buchert2006}, the overall `cosmic equation of state' $w^{\mathcal{D}}_{\rm eff}$ is given by
\begin{equation}\label{eq:29}
  w^{\mathcal{D}}_{\rm eff}=w^{\mathcal{D}}_{\Phi}(1-{\Omega^{\mathcal{D}}_m})\  ,
\end{equation}
where $w^{\mathcal{D}}_{\Phi}$, the constant equation of state for the morphon field, represents the effect of the averaged geometrical degrees of freedom.

\section{Constraints with OHD}
\label{sec:4}

\subsection{The flat FLRW model}
In flat FLRW model, we can have
\begin{equation}\label{eq:30}
  1+z_{\mathcal{D}}=\frac{a_{{\mathcal{D}}_0}}{a_{\mathcal{D}}}\  ,
\end{equation}
where $z_{\mathcal{D}}$, $a_{{\mathcal{D}}_0}$ and $a_{\mathcal{D}}$ are assumed to be identical to $z$, $a_0$ and $a$, respectively.
As a result, Eq. (\ref{eq:18}) becomes
\begin{equation}\label{eq:31}
  H^2_{\mathcal{D}}(z_{\mathcal{D}})=H^2_{{\mathcal{D}}_0}[\Omega^{{\mathcal{D}}_0}_m(1+z_{\mathcal{D}})^{3}+\Omega^{{\mathcal{D}}_0}_X(1+z_{\mathcal{D}})^{-n}],
\end{equation}
where $\Omega^{{\mathcal{D}}_0}_X=1-\Omega^{{\mathcal{D}}_0}_m$, and the unknown parameters are $n$, $\Omega^{{\mathcal{D}}_0}_m$, and $H_{{\mathcal{D}}_0}$. We consider all prior distributions of these parameters to be uniform distributions with ranges of $n$, $\Omega^{{\mathcal{D}}_0}_m$, $H_{{\mathcal{D}}_0}$ from -3 to 3, 0.0 to 0.7, and 50.0 to 90.0, respectively. The constraints on ($n, \Omega^{{\mathcal{D}}_0}_m$) can be obtained by minimizing $\chi^2_H(n, \Omega^{{\mathcal{D}}_0}_m, H_{{\mathcal{D}}_0})$,
\begin{equation}\label{eq:32}
\begin{aligned}
  \chi^2_H(n, \Omega^{{\mathcal{D}}_0}_m, &H_{{\mathcal{D}}_0})=\\
  &\sum_i\frac{[H_{\rm obs}(z_i)-H_{\rm th}(z_i,H_{{\mathcal{D}}_0},n,\Omega^{{\mathcal{D}}_0}_m)]^2}{\sigma^2_{H_i}}\  .
\end{aligned}
\end{equation}
Here we assume that each measurement in $\{H_{\rm obs}(z_i)\}$ is independent. However, we note that the covariance matrix of data is not necessarily diagonal, as discussed in \cite{Yu2013}, and if not, the case will become complicated and should be treated by means of the method mentioned by \cite{Yu2013}. Despite that, if interested in the constraint of $n$ and $\Omega^{{\mathcal{D}}_0}_m$, we could still marginalize $H_{{\mathcal{D}}_0}$ to obtain the probability distribution function of $n$ and $\Omega^{{\mathcal{D}}_0}_m$, i.e., the likelihood function is
\begin{equation}\label{eq:33}
  \mathcal{L}(n,\Omega^{{\mathcal{D}}_0}_m)=\int {\rm d}H_{{\mathcal{D}}_0}P(H_{{\mathcal{D}}_0})e^{-\chi^2_H(n, \Omega^{{\mathcal{D}}_0}_m, H_{{\mathcal{D}}_0})/2} ,
\end{equation}
where $P(H_{{\mathcal{D}}_0})$ is the prior distribution function for the present effective volume Hubble constant. Table \ref{tab:Hubble} \cite{Duan2016} shows all 38 available OHD and reference therein, which includes data obtained by both the differential ages method and the radial Baryon Acoustic Oscillation (BAO) method.
\begin{table}
\centering
\setlength{\tabcolsep}{1mm}{
\begin{tabular}{|lccc|}
\hline
{$z$}   & $H(z)$ & Method & Ref.\\
\hline
$0.0708$   &  $69.0\pm19.68$      &  I    &  Zhang et al. (2014)-\cite{Zhang2014}   \\
    $0.09$       &  $69.0\pm12.0$        &  I    &  Jimenez et al. (2003)-\cite{Jimenez2003}   \\
    $0.12$       &  $68.6\pm26.2$        &  I    &  Zhang et al. (2014)-\cite{Zhang2014}  \\
    $0.17$       &  $83.0\pm8.0$          &  I    &  Simon et al. (2005)-\cite{Simon2005}     \\
    $0.179$     &  $75.0\pm4.0$          &  I    &  Moresco et al. (2012)-\cite{Moresco2012}     \\
    $0.199$     &  $75.0\pm5.0$          &  I    &  Moresco et al. (2012)-\cite{Moresco2012}     \\
    $0.20$         &  $72.9\pm29.6$        &  I    &  Zhang et al. (2014)-\cite{Zhang2014}   \\
    $0.240$     &  $79.69\pm2.65$      &  II   &  Gazta$\tilde{\rm{n}}$aga et al. (2009)-\cite{Gaztanaga2009}   \\
    $0.27$       &  $77.0\pm14.0$        &  I    &    Simon et al. (2005)-\cite{Simon2005}   \\
    $0.28$       &  $88.8\pm36.6$        &  I    &  Zhang et al. (2014)-\cite{Zhang2014}   \\
    $0.35$       &  $84.4\pm7.0$          &  II   &   Xu et al. (2013)-\cite{Xu2013}  \\
    $0.352$     &  $83.0\pm14.0$        &  I    &  Moresco et al. (2012)-\cite{Moresco2012}   \\
    $0.3802$     &  $83.0\pm13.5$        &  I    &  Moresco et al. (2016)-\cite{Moresco2016}   \\
    $0.4$         &  $95\pm17.0$           &  I    &  Simon et al. (2005)-\cite{Simon2005}     \\
    $0.4004$     &  $77.0\pm10.2$        &  I    &  Moresco et al. (2016)-\cite{Moresco2016}   \\
    $0.4247$     &  $87.1\pm11.2$        &  I    &  Moresco et al. (2016)-\cite{Moresco2016}   \\
    $0.43$     &  $86.45\pm3.68$        &  II   &  Gaztanaga et al. (2009)-\cite{Gaztanaga2009}   \\
    $0.44$       & $82.6\pm7.8$           &  II   &  Blake et al. (2012)-\cite{Blake2012}  \\
    $0.4497$     &  $92.8\pm12.9$        &  I    &  Moresco et al. (2016)-\cite{Moresco2016}   \\
    $0.4783$     &  $80.9\pm9.0$        &  I    &  Moresco et al. (2016)-\cite{Moresco2016}   \\
    $0.48$       &  $97.0\pm62.0$        &  I    &  Stern et al. (2010)-\cite{Stern2010}     \\
    $0.57$       &  $92.4\pm4.5$          &  II   &  Samushia et al. (2013)-\cite{Samushia2013}   \\
    $0.593$     &  $104.0\pm13.0$      &  I    &  Moresco et al. (2012)-\cite{Moresco2012}   \\
    $0.6$         &  $87.9\pm6.1$          &  II   &  Blake et al. (2012)-\cite{Blake2012}   \\
    $0.68$       &  $92.0\pm8.0$          &  I    &  Moresco et al. (2012)-\cite{Moresco2012}   \\
    $0.73$       &  $97.3\pm7.0$          &  II   &  Blake et al. (2012)-\cite{Blake2012}  \\
    $0.781$     &  $105.0\pm12.0$      &  I    &  Moresco et al. (2012)-\cite{Moresco2012}   \\
    $0.875$     &  $125.0\pm17.0$      &  I    &  Moresco et al. (2012)-\cite{Moresco2012}   \\
    $0.88$       &  $90.0\pm40.0$        &  I    &  Stern et al. (2010)-\cite{Stern2010}     \\
    $0.9$         &  $117.0\pm23.0$      &  I    &  Simon et al. (2005)-\cite{Simon2005}  \\
    $1.037$     &  $154.0\pm20.0$      &  I    &  Moresco et al. (2012)-\cite{Moresco2012}   \\
    $1.3$         &  $168.0\pm17.0$      &  I    &  Simon et al. (2005)-\cite{Simon2005}     \\
    $1.363$     &  $160.0\pm33.6$      &  I    &  Moresco (2015)-\cite{Moresco2015}  \\
    $1.43$       &  $177.0\pm18.0$      &  I    &  Simon et al. (2005)-\cite{Simon2005}     \\
    $1.53$       &  $140.0\pm14.0$      &  I    &  Simon et al. (2005)-\cite{Simon2005}     \\
    $1.75$       &  $202.0\pm40.0$      &  I    &  Simon et al. (2005)-\cite{Simon2005}     \\
    $1.965$     &  $186.5\pm50.4$      &  I    &   Moresco (2015)-\cite{Moresco2015}  \\
    $2.34$       &  $222.0\pm7.0$        &  II   &  Delubac et al. (2015)-\cite{Delubac2015}   \\
\hline
\end{tabular}}
\caption{\label{tab:Hubble} The current available OHD dataset. The method I is the differential ages method, and II represents the radial Baryon Acoustic Oscillation (BAO) method. $H(z)$ is in unit of ${\rm km \ s^{-1} \ Mpc^{-1}}$ here.}
\end{table}
\begin{figure*}
\centering
\includegraphics[width=0.63\textwidth]{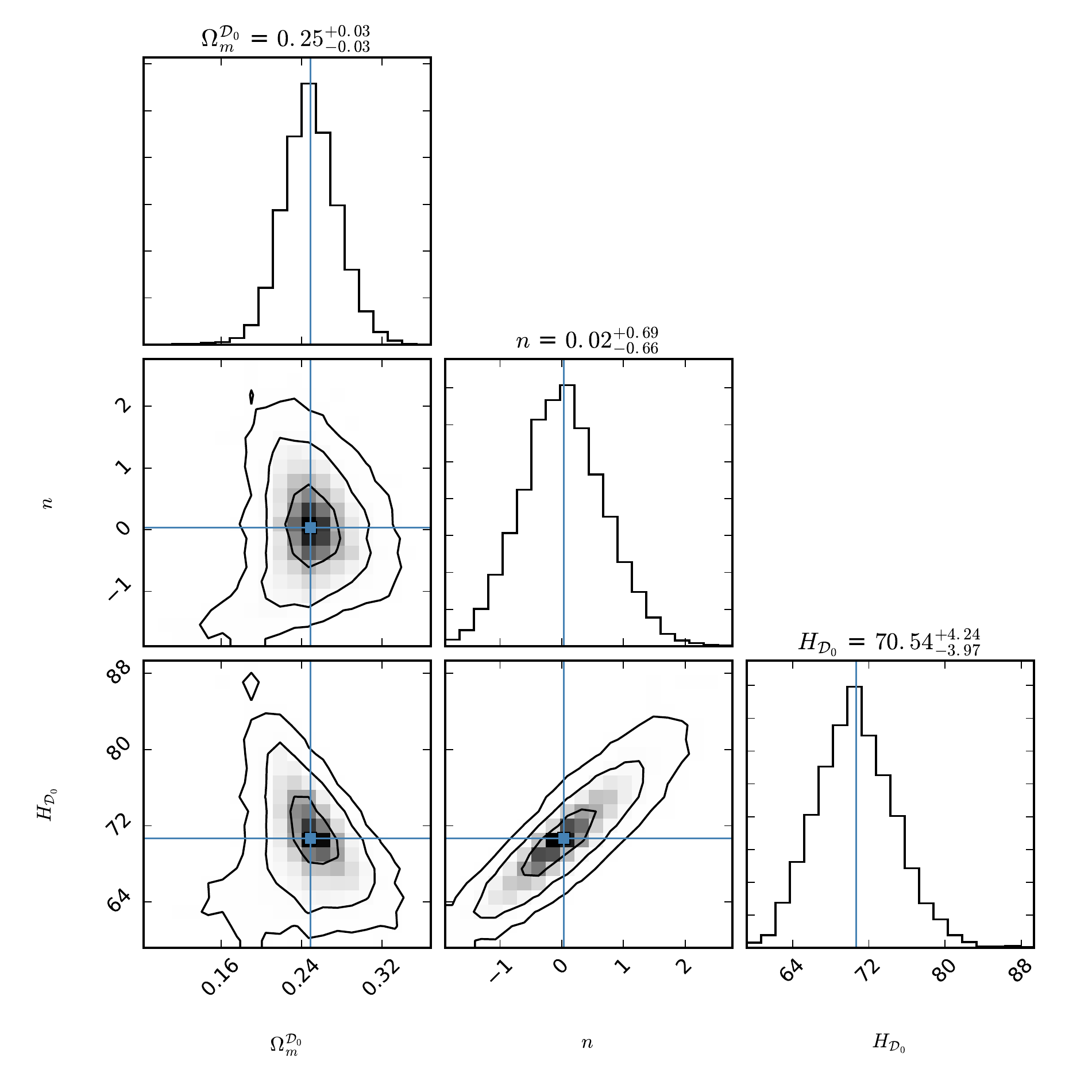}
\caption{The 1$\sigma$, 2$\sigma$ and 3$\sigma$ confidence regions of the effective parameters $n$, $\Omega^{{\mathcal{D}}_0}_m$, and $H_{{\mathcal{D}}_0}$ for the backreaction model, along with their own probability density fuction.}\label{fig:2}
\end{figure*}

The confidence regions are shown in Fig. \ref{fig:2}, where the best fits are $\Omega^{{\mathcal{D}}_0}_m = 0.25^{+0.03}_{-0.03}$, $n = 0.02^{+0.69}_{-0.66}$, and $H_{\mathcal{D}_0} = 70.54^{+4.24}_{-3.97}\ {\rm km \ s^{-1} \ Mpc^{-1}}$. As opposed to Fig. 2 of \cite{Larena2009}, the degeneracy direction \cite{Teng2016} of the contour ($n, \Omega^{{\mathcal{D}}_0}_m$) is different in our Fig. \ref{fig:2}. In this paper, the best-fit values of $\Omega^{{\mathcal{D}}_0}_m=0.25$ and $n=0.03$, while in \cite{Larena2009}, $\Omega^{{\mathcal{D}}_0}_m = 0.26$ and $n = 0.24$ were given for the flat FLRW model. The best-fits of $\Omega^{{\mathcal{D}}_0}_m$ are almost the same, however, the values of $n$ are variant, the reason of which may be the lack of precision caused by the insufficient amount of OHD. Nevertheless, as for best-fit of $\Omega^{{\mathcal{D}}_0}_m$, in comparison with \cite{Astier2006}, $\Omega_m = 0.263^{+0.042}_{-0.042}(1\sigma\ \rm stat)\\^{+0.032}_{-0.032}(\rm sys)$ for a flat $\Lambda$CDM model, and with \cite{Perlmutter1999}, $\Omega^{flat}_m = 0.28^{+0.09}_{-0.08}(1\sigma\ \rm stat)^{+0.05}_{-0.04}(\rm sys)$ for a flat cosmology, the proportions of matter density are not much of differences.

\begin{figure*}
\centering
\includegraphics[width=0.63\textwidth]{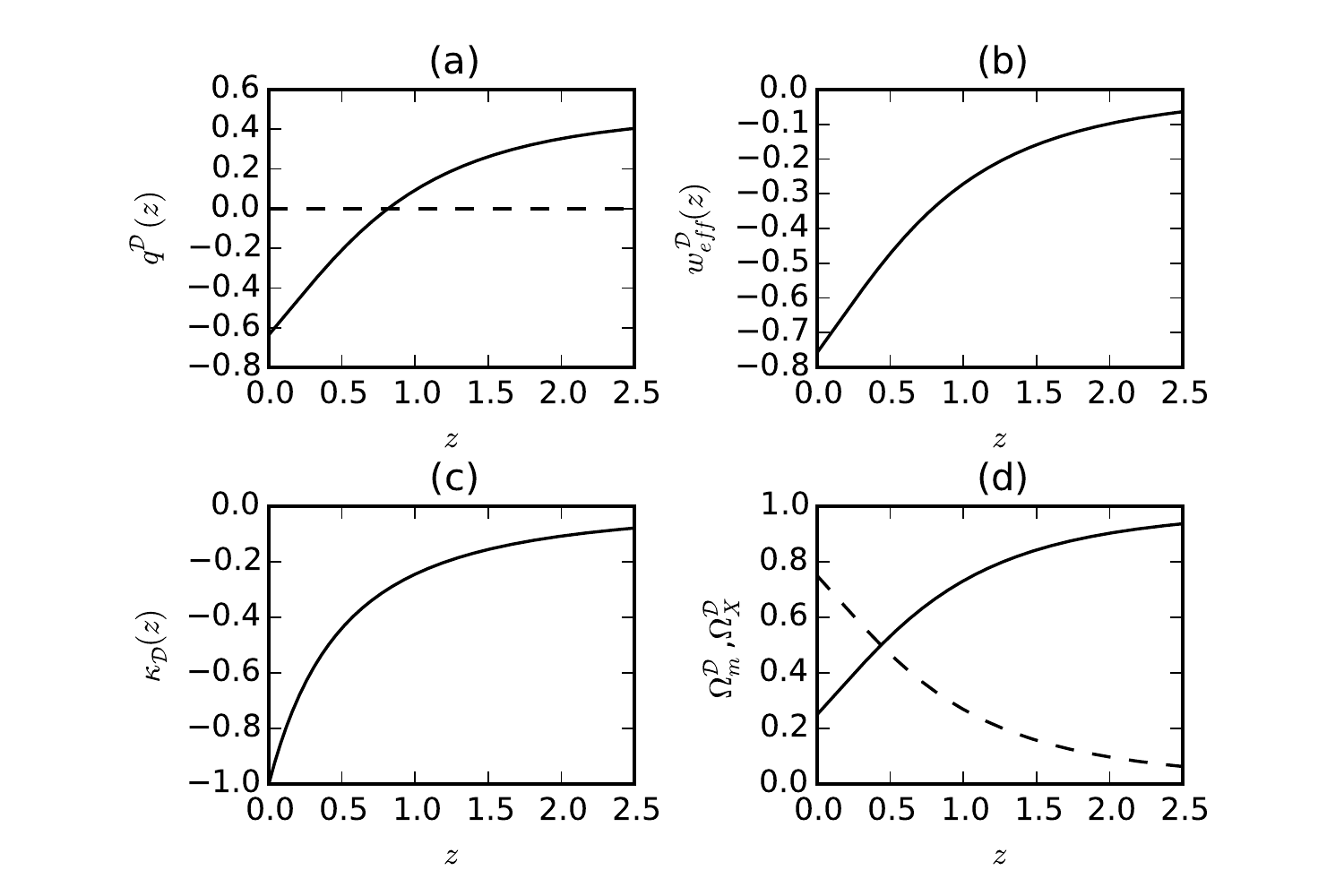}
\caption{(a) The evolution of $q^{\mathcal{D}}$ with best-fit values of $n$ = 0.03, and $\Omega^{{\mathcal{D}}_0}_m$ = 0.25; (b) The evolution of $w^{\mathcal{D}}_{\rm eff}(z)$ with the same best-fit values as in (a); (c) The relation between $\kappa_{\mathcal{D}}(z)$ and $z$ with the same best-fit values as in (a); (d) The evolution of $\Omega^{\mathcal{D}}_m$ (solid line) and $\Omega^{\mathcal{D}}_X$ (dashed line) with the same best fits as in (a).}\label{fig:3}
\end{figure*}

The evolutions of $\kappa_{\mathcal{D}}(z)$ and the dimensionless averaged cosmological parameters are illustrated in Fig. \ref{fig:3} (c) and (d), respectively, with the best fits of $n$ = 0.03, and $\Omega^{{\mathcal{D}}_0}_m$ = 0.25. The former obviously evolves from 0 to -1, which is biased due to the prescription. Furthermore, we substitute the best-fits into the effective volume deceleration parameter $q^{\mathcal{D}}$
\begin{equation}\label{eq:34}
q^{\mathcal{D}}=-\frac{n+2}{2}+\frac{n+3}{2}\Omega^{\mathcal{D}}_m\  ,
\end{equation}
where
\begin{equation}
\Omega^{\mathcal{D}}_m=\frac{\Omega^{{\mathcal{D}}_0}_m}{\Omega^{{\mathcal{D}}_0}_m+(1-\Omega^{{\mathcal{D}}_0}_m)(1+z)^{-(n+3)}}\  .\nonumber
\end{equation}
Consequently, Fig. \ref{fig:3} (a) illustrates how the volume deceleration parameter $q^{\mathcal{D}}$ evolves over redshift $z$ with best-fit values of $n$ and $\Omega^{{\mathcal{D}}_0}_m$. The transition redshift $z_t$, when the universe transited from a deceleration to accelerated expansion phase, is around 0.815, and the present value of the volume deceleration parameter $q^{{\mathcal{D}}_0}$ is about -0.636. While the transition value in \cite{Riess2004} is constrained to be $z_t = 0.46 \pm 0.13$, and in \cite{Santos2016} is $z_t = 0.68^{+0.10}_{-0.09}$ with the present value of the deceleration parameter $q_0 = -0.48^{+0.11}_{-0.14}$, we find that our values of $z_t$ and $q^{{\mathcal{D}}_0}$ are both higher than them. Of course, since the flat FLRW metric does not fit averaged model, the result is predetermined. As shown in Fig. \ref{fig:3} (b), the present value $w^{{\mathcal{D}}_0}_{\rm eff}$ is approximately -0.758.

\subsection{The template metric model}

\begin{figure*}[htbp]
\centering
\includegraphics[width=0.63\textwidth]{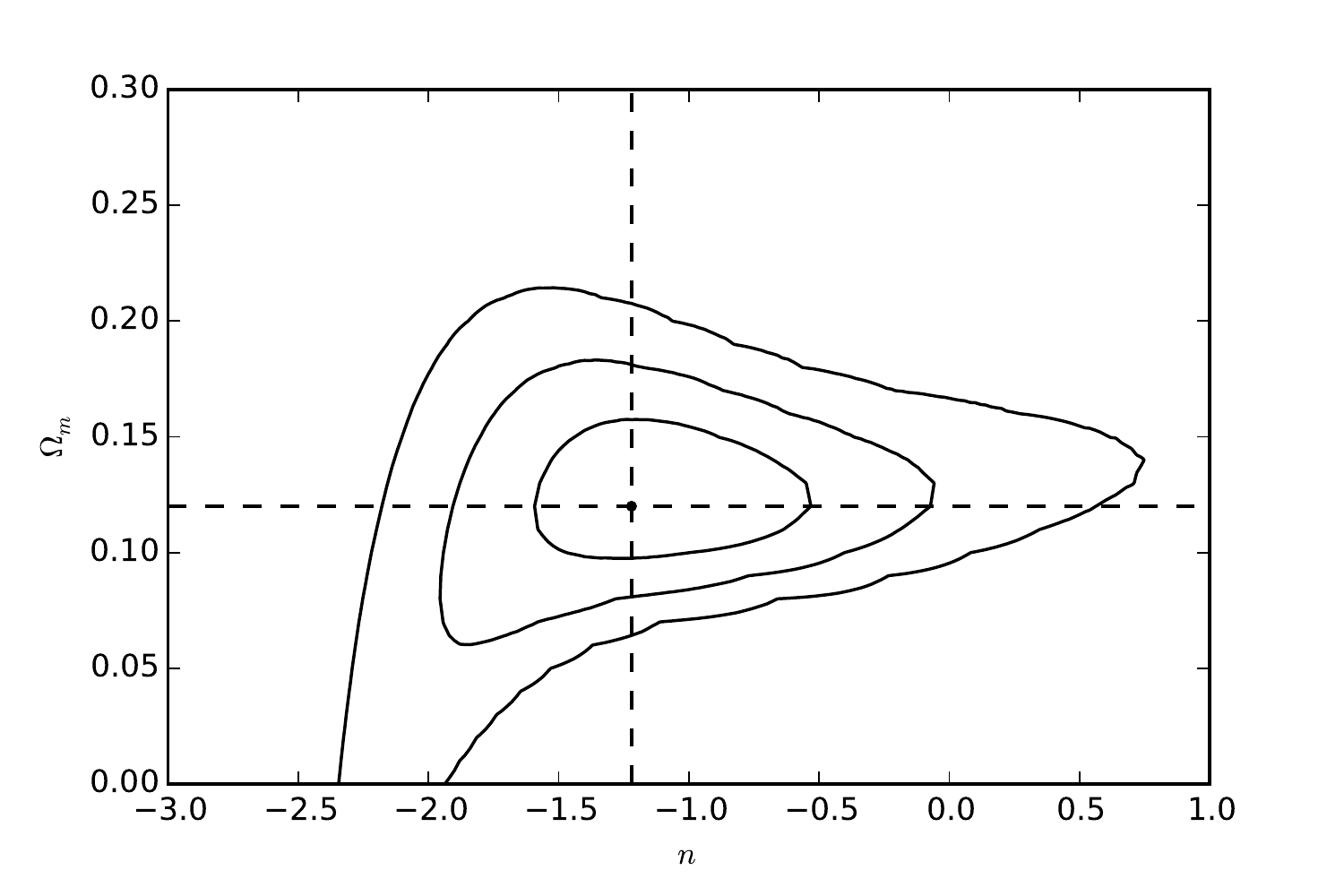}
\caption{The 1$\sigma$, 2$\sigma$ and 3$\sigma$ confidence regions of the effective parameters $n$, $\Omega^{{\mathcal{D}}_0}_m$, marginalizing $H_{{\mathcal{D}}_0}$ over the interval [50, 90], where the best-fits are $n=-1.22$, and ${{\Omega}_{m}^{\mathcal{D}_{0}}}=0.12$.}\label{fig:4}
\end{figure*}

\begin{figure*}[htbp]
\centering
\includegraphics[width=0.63\textwidth]{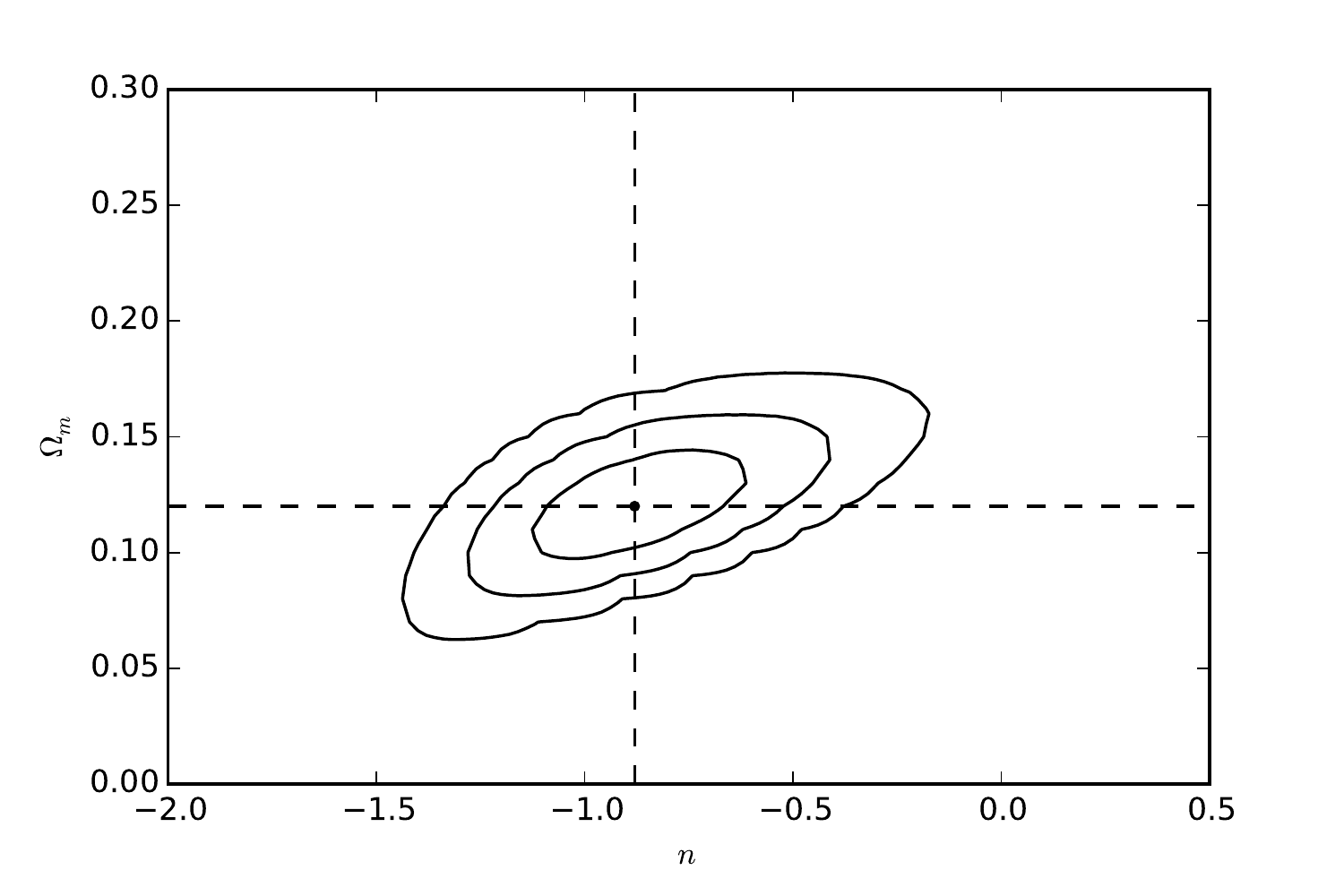}
\caption{The 1$\sigma$, 2$\sigma$ and 3$\sigma$ confidence regions of the effective parameters $n$, $\Omega^{{\mathcal{D}}_0}_m$ with Gaussian prior of $H_{{\mathcal{D}}_0}=69.32\pm0.80\  \rm km \ s^{-1} \ Mpc^{-1}$, where the best-fits are $n=-0.88$, and ${{\Omega}_{m}^{\mathcal{D}_{0}}}=0.12$.}\label{fig:5}
\end{figure*}

\begin{figure*}[htbp]
\centering
\includegraphics[width=0.63\textwidth]{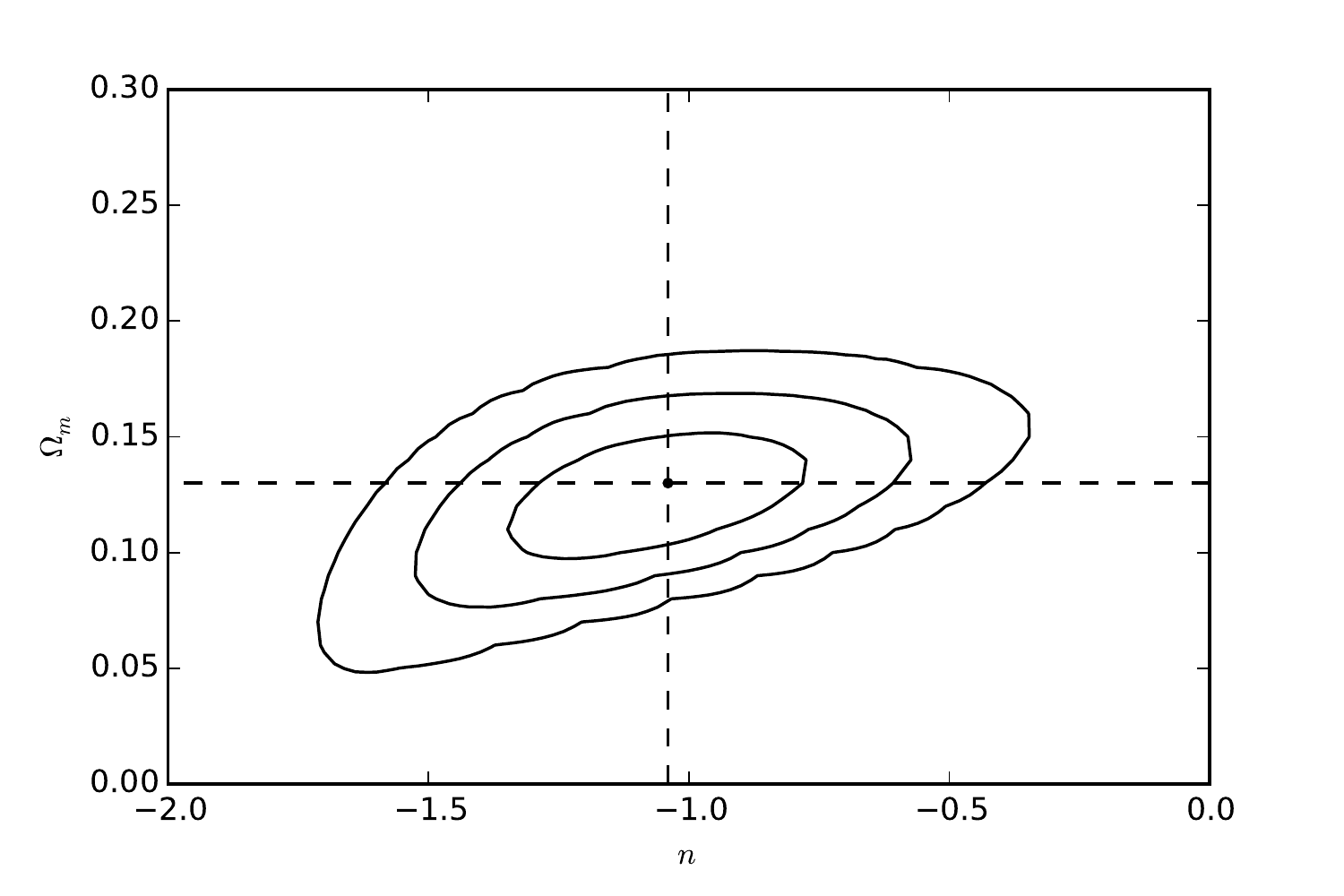}
\caption{Same as Fig. \ref{fig:5}, but with Gaussian prior of $H_{{\mathcal{D}}_0}=67.3\pm1.2\  \rm km \ s^{-1} \ Mpc^{-1}$, where the best-fits are $n=-1.04$, and ${{\Omega}_{m}^{\mathcal{D}_{0}}}=0.13$.}\label{fig:6}
\end{figure*}

\begin{figure*}[htbp]
\centering
\includegraphics[width=0.63\textwidth]{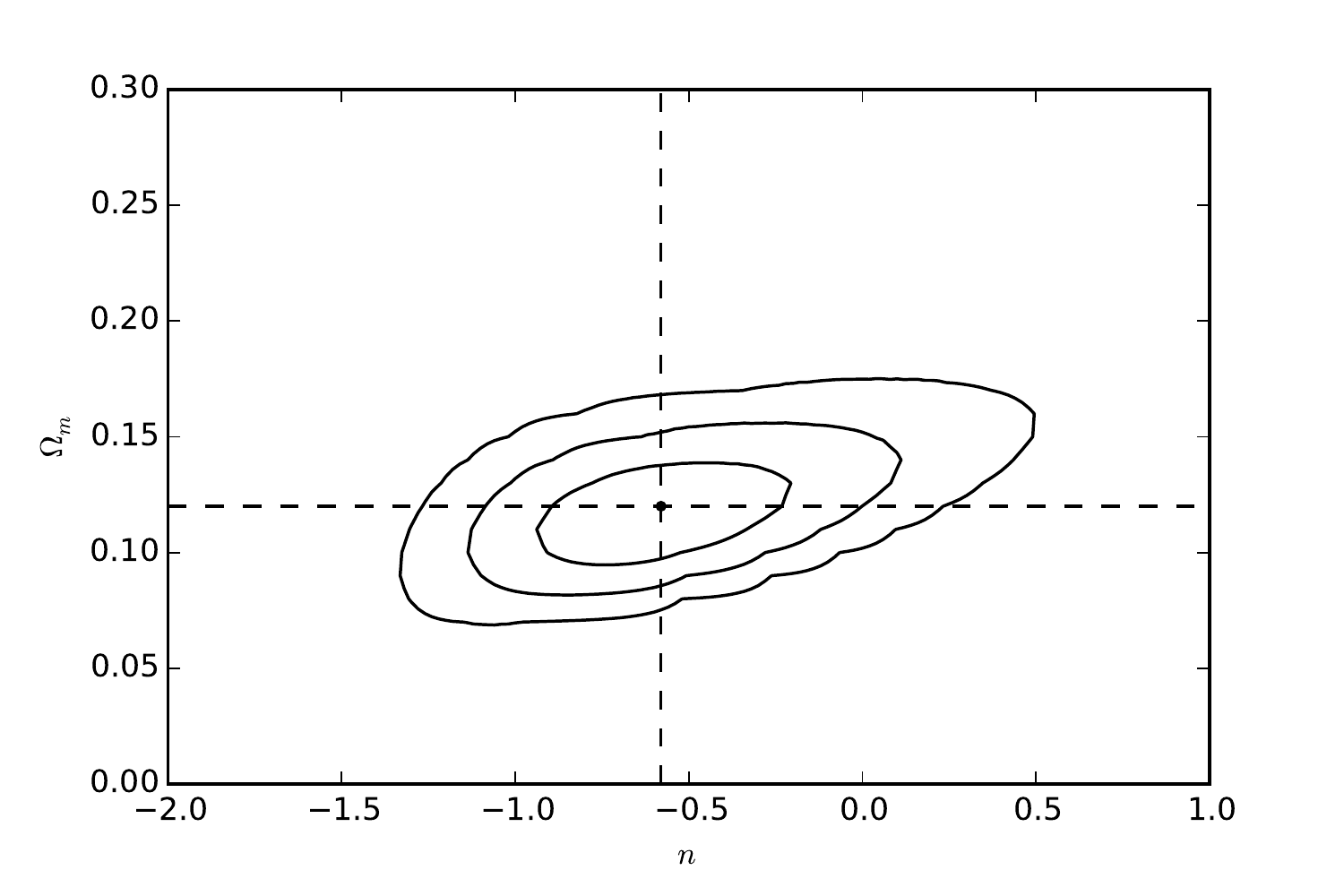}
\caption{Same as Fig. \ref{fig:5}, but with Gaussian prior of $H_{{\mathcal{D}}_0}=73.24\pm1.74\  \rm km \ s^{-1} \ Mpc^{-1}$, where the best-fits are $n=-0.58$, and ${{\Omega}_{m}^{\mathcal{D}_{0}}}=0.12$.}\label{fig:7}
\end{figure*}

In the template metric context, as mentioned above, we cannot directly constrain parameters with the expression of Eq. (\ref{eq:18}). However, by using Runge-Kutta method to solve Eq. (\ref{eq:28}), Eq. (\ref{eq:25}) and subsequently Eq. (\ref{eq:24}) with given values of $n$ and $\Omega^{{\mathcal{D}}_0}_m$, we can form a one-to-one correspondence of $a_{\mathcal{D}}$ and $z_{\mathcal{D}}$, and then the constraint on parameters of Eq. (\ref{eq:18}) with OHD could be performed. Thus, we can rewrite Eq. (\ref{eq:18}) as follows
\begin{equation}\label{eq:35}
\begin{aligned}
  &H_{\mathcal{D}}(a_{\mathcal{D}})=H_{{\mathcal{D}}_0}\sqrt{\Omega^{{\mathcal{D}}_0}_m{a^{-3}_{\mathcal{D}}}+(1-\Omega^{{\mathcal{D}}_0}_m){a^n_{\mathcal{D}}}}\\
  &\Leftrightarrow H_{\mathcal{D}}(z_{\mathcal{D}})=H_{{\mathcal{D}}_0}E(z_{\mathcal{D}}, n, \Omega^{{\mathcal{D}}_0}_m)\  .
\end{aligned}
\end{equation}
After marginalizing the likelihood function over $H_{{\mathcal{D}}_0}$, we can obtain parameter constraints in ($n$, $\Omega^{{\mathcal{D}}_0}_m$) subspace. As Ma et. al. \cite{Ma2011} stated, with top-hat prior of $H_{{\mathcal{D}}_0}$ over the interval $[x, y]$, the posterior probability density function (PDF) of parameters given the dataset \{$H_i$\} by Bayes' theorem reads
\begin{equation}\label{eq:36}
  P(n,\Omega^{{\mathcal{D}}_0}_m\mid\{H_i\})=\frac{U(x,C,D)-U(y,C,D)}{\sqrt{C}}{\rm exp}(\frac{D^2}{C})
\end{equation}
where
\begin{equation*}
  C=\sum_i\frac{E^2(z_i;n,\Omega^{{\mathcal{D}}_0}_m)}{2\sigma^2_i},\ D=\sum_i\frac{E(z_i;n,\Omega^{{\mathcal{D}}_0}_m)H_i}{2\sigma^2_i},
\end{equation*}
and
\begin{equation*}
  U(x,\alpha,\beta)={\rm erf}(\frac{\beta-x\alpha}{\sqrt{\alpha}}),
\end{equation*}
where $x=50.0$, $y=90.0$, and erf represents the error \\function. In the process here, we utilize so called mesh-grid\\
 method to scan spots of ($n, \Omega^{{\mathcal{D}}_0}_m$) subspace with the range of $n$, and $\Omega^{{\mathcal{D}}_0}_m$ to be [-3, 3] and [0.0, 0.7], respectively. Note that the range of $n$ is not randomly chosen. Since at first, we are not sure for the specific region, due to the symmetrical purpose and ranging from large to small, we gradually narrow down the regions into the possible region that is enough for all the constraining cases. Eventually, as shown in Fig. \ref{fig:4}, we attain the constraints with $n=-1.22^{+0.68}_{-0.40}$, and ${{\Omega}_{m}^{\mathcal{D}_{0}}}=0.12^{+0.04}_{-0.02}$. Apparently, our results are quite different from Larena's results of $\Omega^{{\mathcal{D}}_0}_m = 0.397$ and $n = 0.5$ for averaged model. Since these are all based on the power law ansatz of $a_{\mathcal{D}}$, the issues might be the wrong prescription of $\kappa_{\mathcal{D}}$, the chosen prior of $H_{{\mathcal{D}}_0}$, or the lack of amount for OHD.
To test the probability of the second issue, we select three different Gaussian prior distributions of $H_{{\mathcal{D}}_0}$. Thus, as described in \cite{Ma2011}, the posterior PDF of parameters becomes
\begin{equation}\label{eq:37}
  P(n,\Omega^{{\mathcal{D}}_0}_m\mid\{H_i\})=\frac{1}{\sqrt{A}}[{\rm erf}(\frac{B}{\sqrt{A}})+1]{\rm exp}(\frac{B^2}{A}),
\end{equation}
where
\begin{equation*}
  A=\frac{1}{2\sigma^2_H}\sum_i\frac{E^2(z_i;n,\Omega^{{\mathcal{D}}_0}_m)}{2\sigma^2_i},
\end{equation*}
\begin{equation*}
  B=\frac{\mu_H}{2\sigma^2_H}\sum_i\frac{E^2(z_i;n,\Omega^{{\mathcal{D}}_0}_m)H_i}{2\sigma^2_i},
\end{equation*}
where $\mu_H$ and $\sigma^2_H$ denote prior expectation and deviation of $H_{{\mathcal{D}}_0}$, respectively.
First, we make use of $H_{{\mathcal{D}}_0}=69.32\pm0.80\  \rm km \ s^{-1} \ Mpc^{-1}$ \cite{Bennett2013} to obtain the constraints. As a result, Fig. \ref{fig:5} illustrates the confidence regions with 1$\sigma$ constraints $n=-0.88^{+0.26}_{-0.23}$, and ${{\Omega}_{m}^{\mathcal{D}_{0}}}=0.12^{+0.03}_{-0.03}$. Second, as shown in Fig. \ref{fig:6}, we acquire the constraints $n=-1.04^{+0.27}_{-0.31}$ and ${{\Omega}_{m}^{\mathcal{D}_{0}}}=0.13^{+0.02}_{-0.03}$ with $H_{{\mathcal{D}}_0}=67.3\pm1.2\  \rm km \ s^{-1} \ Mpc^{-1}$ \cite{Planck2014}. Last, we attain the constraints with Gaussian prior of $H_{{\mathcal{D}}_0}=73.24\pm1.74\  \rm km \ s^{-1} \ Mpc^{-1}$ \cite{Riess2016}, as depicted in Fig. \ref{fig:7}, $n=-0.58^{+0.36}_{-0.34}$ and ${{\Omega}_{m}^{\mathcal{D}_{0}}}=0.12^{+0.02}_{-0.03}$. As a result, we find that although three Gaussian priors lead to different best-fit values of $n$, the best-fits of ${{\Omega}_{m}^{\mathcal{D}_{0}}}$ are compatible with the result of top-hat prior.

\begin{figure*}[htbp]
\centering
\includegraphics[width=0.65\textwidth]{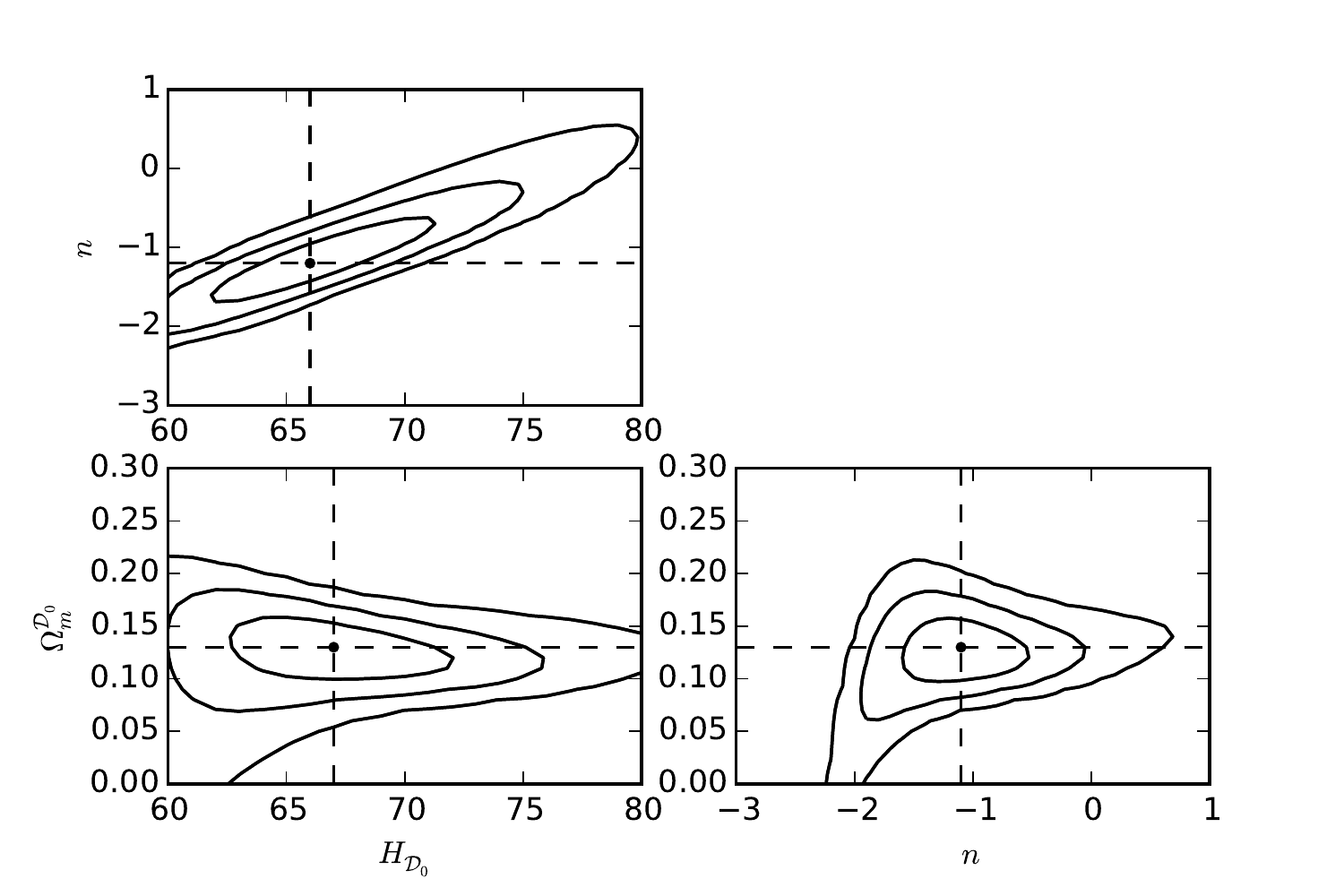}
\caption{The likelihood contour without marginalization of the parameters, where the best-fit pairs are $(n, {{H}_{\mathcal{D}_{0}}})=(-1.2, 66)$, $({{\Omega}_{m}^{\mathcal{D}_{0}}}, {{H}_{\mathcal{D}_{0}}})=(0.13, 67)$, and $(n, {{\Omega}_{m}^{\mathcal{D}_{0}}})=(-1.1, 0.13)$.}\label{fig:8}
\end{figure*}

Furthermore, we constrain the model without margi-\\nalization of the Hubble constant. The absolute best-fit results are plotted in Fig. \ref{fig:8}, where the 1$\sigma$ best-fit values are ($n=-1.2^{+0.61}_{-0.58}, {{H}_{\mathcal{D}_{0}}}=66^{+5.3}_{-4.2}\ \rm km\ s^{-1}\ Mpc^{-1}$), (${{\Omega}_{m}^{\mathcal{D}_{0}}}=0.13\pm0.03, {{H}_{\mathcal{D}_{0}}}=67^{+5.1}_{-4.4}\ \rm km\ s^{-1}\ Mpc^{-1}$), and ($n=-1.1\\^{+0.58}_{-0.50},{{\Omega}_{m}^{\mathcal{D}_{0}}}=0.13\pm0.03$). The best-fits results, $n=-1.1$ and ${{\Omega}_{m}^{\mathcal{D}_{0}}}=0.13$, are consistent with both the ones with the Gaussian prior of $H_{{\mathcal{D}}_0}=67.3\pm1.2\  \rm km \ s^{-1}\ Mpc^{-1}$ and the ones with top-hat prior of $H_{{\mathcal{D}}_0}$, and also in contrary with the absolute constraint results of Larena et. al., i.e., $n=0.12$ and ${{\Omega}_{m}^{\mathcal{D}_{0}}}=0.38$. Therefore, the prior issue can be ruled out.


Since the same set of data shared by FLRW case results in a reasonable conclusion, the lack of amount for OHD can also be neglected. Therefore, the prescription of $\kappa_{\mathcal{D}}$ should be modified, or some other scenarios should be introduced. In fact, based on subsection 4.3 of \cite{Larena2009}, the most probable value of $n-m$ should be -1, i.e., the prescription should be modified as follows,
\begin{equation}\label{eq:38}
  \langle {\mathcal{R}}\rangle_{\mathcal{D}}=\frac{\kappa_{\mathcal{D}}(t)|\langle {\mathcal{R}}\rangle_{{\mathcal{D}}_0}|a_{{\mathcal{D}}_0}}{a_{\mathcal{D}}(t)}\  .
\end{equation}
This form can also guarantee that $\kappa_{\mathcal{D}}$ inherits the sign of $\langle {\mathcal{R}}\rangle_{\mathcal{D}}$, as the sign of $a_{\mathcal{D}}$ is positive. Moreover, since the backreaction is considered, $\kappa_{\mathcal{D}}$ can be related to both ${\mathcal{Q}}_{\mathcal{D}}$ and $\langle {\mathcal{R}}\rangle_{\mathcal{D}}$. In this context, the following form can be evaluated
\begin{equation}\label{eq:39}
  \langle {\mathcal{R}}\rangle_{\mathcal{D}}+{\mathcal{Q}}_{\mathcal{D}}=\frac{\kappa_{\mathcal{D}}(t)|\langle {\mathcal{R}}\rangle_{{\mathcal{D}}_0}+{\mathcal{Q}}_{\mathcal{D}_0}|a_{{\mathcal{D}}_0}}{a_{\mathcal{D}}(t)}\  .
\end{equation}
However, because of the coupling between ${\mathcal{Q}}_{\mathcal{D}}$ and $\langle {\mathcal{R}}\rangle_{\mathcal{D}}$, there is not much difference for the above two forms. But if ${\mathcal{Q}}_{\mathcal{D}}$ and $\langle {\mathcal{R}}\rangle_{\mathcal{D}}$ are not following the scaling solutions, then the difference would be significant. Note that even if we choose the $n\cong m$ case, the form of $\kappa_{\mathcal{D}}$ should be
\begin{equation}\label{eq:40}
  \langle {\mathcal{R}}\rangle_{\mathcal{D}}=\kappa_{\mathcal{D}}(t)|\langle {\mathcal{R}}\rangle_{{\mathcal{D}}_0}|\  ,
\end{equation}
instead of Eq. (\ref{eq:22}). Besides, in order to compensate the fact that $\kappa_{\mathcal{D}}$ can only be changed from zero to negative numbers, we can add a positive constant to the expression of $\kappa_{\mathcal{D}}$.

Our results show that it demands lower values of $\Omega^{\mathcal{D}_0}_m$ for the models to be compatible with data, and on the contrary with Larena's conclusion, a larger amount of backreaction is required to account for effective geometry. As mentioned in \cite{Larena2009}, a DE model in FLRW  context with $n=-1$ is compatible with the data at 1$\sigma$ for $\Omega^{\mathcal{D}_0}_m \sim 0.1$, and as calculated in \cite{Li2007} and \cite{Li2008}, the leading perturbative model ($n=-1$) is marginally at 1$\sigma$ for $\Omega^{\mathcal{D}_0}_m \sim 0.3$. As expected, purely perturbative estimate of backreaction could not provide sufficient geometrical effect to account for observations. What is not expected is that the values of $\Omega^{\mathcal{D}_0}_m$ is higher or lower compared to the standard DE models with a FLRW geometry. The following subsection will explore the effective deceleration parameter $q^{\mathcal{D}}$ evolves over $a_{\mathcal{D}}$ in many cases, in order to pinpoint the hinge of the issue.

\subsection{Testing the effective deceleration parameter}
\begin{figure*}[htbp]
\centering
\includegraphics[width=0.65\textwidth]{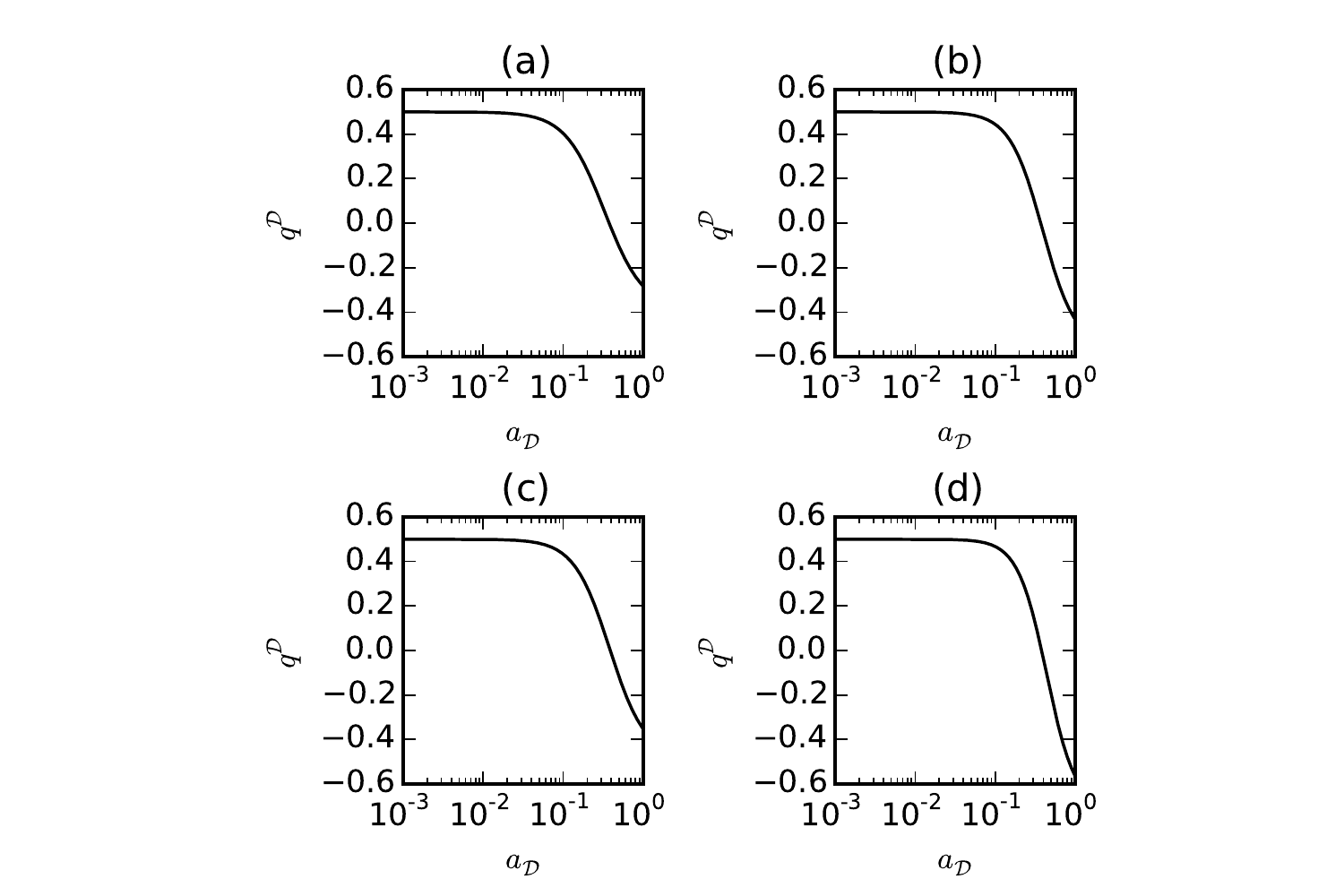}
\caption{The evolutions of $q^{\mathcal{D}}$ over effective scale factor $a_{\mathcal{D}}$ with best-fit values of (a) $n$ = -1.22, $\Omega^{{\mathcal{D}}_0}_m$ = 0.12, (b) $n$ = -0.88, $\Omega^{{\mathcal{D}}_0}_m$ = 0.12, (c) $n$ = -1.04, $\Omega^{{\mathcal{D}}_0}_m$ = 0.13, and (d) $n$ = -0.58, $\Omega^{{\mathcal{D}}_0}_m$ = 0.12, respectively.}\label{fig:9}
\end{figure*}

\begin{figure*}[htbp]
\centering
\includegraphics[width=0.65\textwidth]{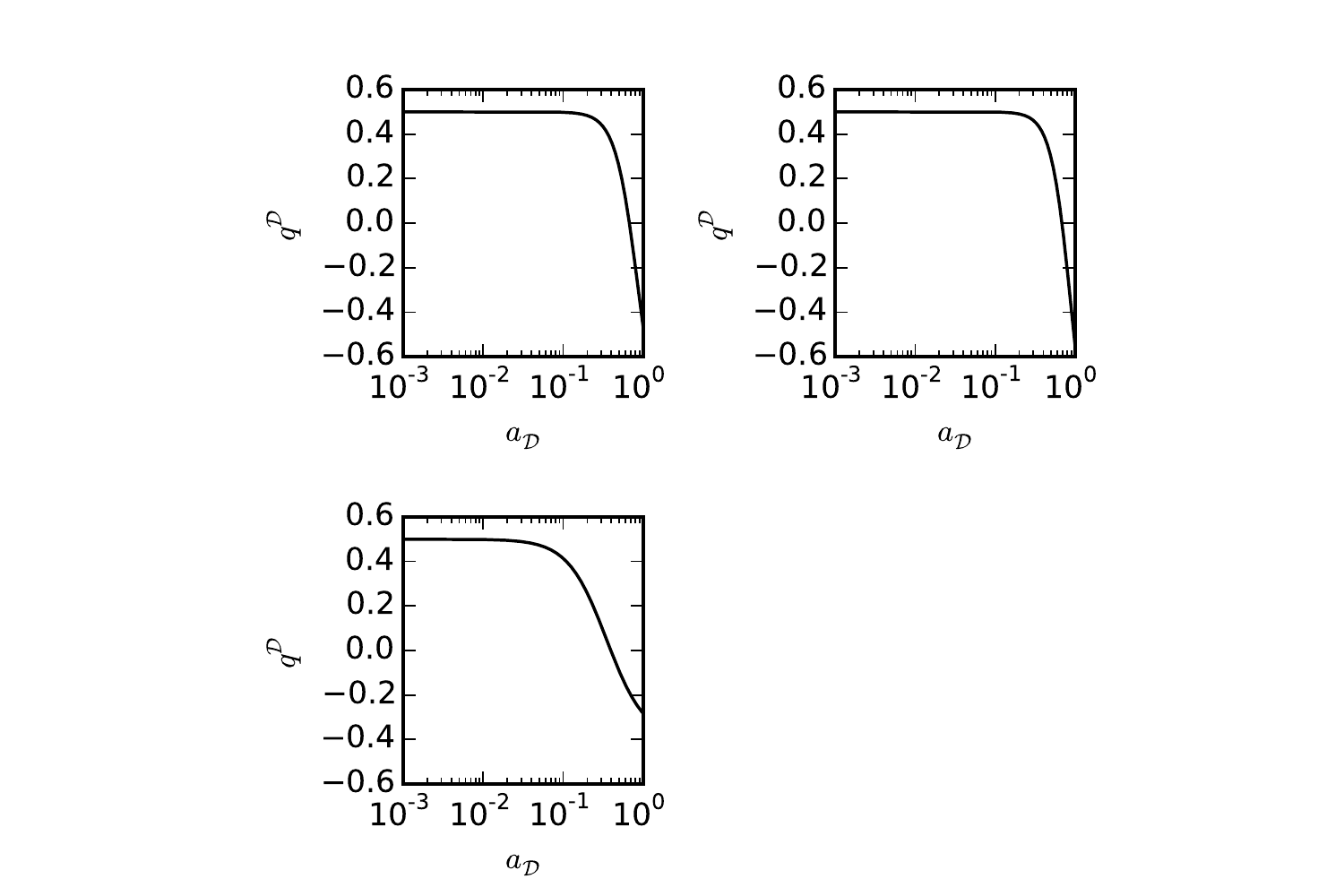}
\caption{Same as Fig. \ref{fig:9}, but for best-fit values of (a) $n$ = 0.12, $\Omega^{{\mathcal{D}}_0}_m$ = 0.38, (b) $n$ = 0.5, $\Omega^{{\mathcal{D}}_0}_m$ = 0.397, and (c) $n$ = -1.1, $\Omega^{{\mathcal{D}}_0}_m$ = 0.13.}\label{fig:10}
\end{figure*}

Fig. \ref{fig:9} shows the evolutions of $q^{\mathcal{D}}$ over effective scale factor $a_{\mathcal{D}}$ with our best-fit values. We can gain the information that in each case $q^{\mathcal{D}}$ tends to 0.5 as $a_{\mathcal{D}}$ becomes smaller, and they all have sufficient backreaction to meet the observations, at leat in this perspective, which indicates that the observational data do not disfavour the constraints. However, we also illustrate the same evolutions by using the absolute and the marginalized best-fits of Larena et. al. and our absolute ones, as shown in Fig. \ref{fig:10}, where the same conclusions are found. The only differences lie in the different turning points for the slow evolutions and present values of $q^{\mathcal{D}_0}$. The larger $n$ becomes, the earlier the evolutionary curves change from fast to slow. This further favours our doubt about the prescription of $\kappa_{\mathcal{D}}$.

\section{Conclusions and discussions}
\label{sec:5}

In this paper, we delve the backreaction model of dust cosmology with both FLRW metric and smoothed template metric to constrain parameters with observational Hubble parameter data (OHD), the purpose of which is to explore the generic properties of a backreaction model for explaining the observations of the Universe. Unlike the work \cite{Chiesa2014}, first, in the FLRW model, we constrain two of three parameters with MCMC method by marginalizing the likelihood function over the rest one parameter, and obtain the best-fits: $\Omega^{{\mathcal{D}}_0}_m = 0.25^{+0.03}_{-0.03}$, $n = 0.02^{+0.69}_{-0.66}$, and $H_{\mathcal{D}_0} = 70.54^{+4.24}_{-3.97}\ \rm km\\
\ s^{-1} \ Mpc^{-1}$. We employ these best-fit values of $n$ and $\Omega^{{\mathcal{D}}_0}_m$ to study the evolutions of $q^{\mathcal{D}}$, $w^{\mathcal{D}}_{\rm eff}$, $\kappa_{\mathcal{D}}$, and effective density parameters. The results compared with other models are slightly biased, which is natural as for the inconsistency between FLRW geometry and averaged model. Second, with template metric and the specific method for computing the observables along null geodesic, we choose a top-hat prior, i.e., uniform distribution of $H_{{\mathcal{D}}_0}$ to be marginalized, in order to attain the posterior PDF of parameters. By making use of classical mesh-grid method, we plot the likelihood contour in the subspace of $(n,\Omega_m^{\mathcal{D}_0})$, and obtain the best-fit values, which are $n=-1.22^{+0.68}_{-0.41}$ and ${{\Omega}_{m}^{\mathcal{D}_{0}}}=0.12^{+0.04}_{-0.02}$. The value of ${{\Omega}_{m}^{\mathcal{D}_{0}}}$ is considerably small in comparison with the one of Larena et. al.\cite{Larena2009}.

Our results show that it demands lower values of $\Omega^{\mathcal{D}_0}_m$ for the models to be compatible with data, which means that on the contrary with Larena's conclusion, a larger amount of backreaction is required to account for effective geometry. The reasons for this discrepancy may be the wrong prescription of $\kappa_{\mathcal{D}}$, the chosen prior of $H_{{\mathcal{D}}_0}$, or the lack of amount for OHD. To test the probability of the second reason, we select three different Gaussian prior distributions of $H_{{\mathcal{D}}_0}$, where the three sets of best-fits are: $n=-0.88^{+0.26}_{-0.23}$, and ${{\Omega}_{m}^{\mathcal{D}_{0}}}=0.12^{+0.03}_{-0.03}$, for $H_{{\mathcal{D}}_0}=69.32\pm0.80\  \rm km \ s^{-1} \ Mpc^{-1}$ \cite{Bennett2013}; $n=-1.04^{+0.27}_{-0.31}$, and ${{\Omega}_{m}^{\mathcal{D}_{0}}}=0.13^{+0.02}_{-0.03}$ with $H_{{\mathcal{D}}_0}=67.3\pm1.2\  \rm km \ s^{-1} \ Mpc^{-1}$ \cite{Planck2014}; $n=-0.58^{+0.36}_{-0.34}$, and ${{\Omega}_{m}^{\mathcal{D}_{0}}}=0.12^{+0.02}_{-0.03}$, related to $H_{{\mathcal{D}}_0}=73.24\pm1.74\  \rm km \ s^{-1} \ Mpc^{-1}$ \cite{Riess2016}. As a result, we find out that although three Gaussian priors lead to different best-fit values of $n$, the best-fits of ${{\Omega}_{m}^{\mathcal{D}_{0}}}$ are still compatible with the result of top-hat prior. In addition, we also constrain the parameters without marginalization of any parameter, and obtain the best-fit values: ($n=-1.2^{+0.61}_{-0.58}, {{H}_{\mathcal{D}_{0}}}=66^{+5.3}_{-4.2}\ \rm km\ s^{-1}\ Mpc^{-1}$), (${{\Omega}_{m}^{\mathcal{D}_{0}}}=0.13\pm0.03, {{H}_{\mathcal{D}_{0}}}=67^{+5.1}_{-4.4}\ \rm km\ s^{-1}\ Mpc^{-1}$), and ($n=-1.1^{+0.58}_{-0.50},{{\Omega}_{m}^{\mathcal{D}_{0}}}=0.13\pm0.03$). The best-fits results, $n=-1.1$ and ${{\Omega}_{m}^{\mathcal{D}_{0}}}=0.13$, are consistent with both the ones with the Gaussian prior of $H_{{\mathcal{D}}_0}=67.3\pm1.2\  \rm km \ s^{-1}\\
\ Mpc^{-1}$ and the ones with top-hat prior of $H_{{\mathcal{D}}_0}$, and also in contrary with the absolute constraint results of \cite{Larena2009}, i.e., $n=0.12$ and ${{\Omega}_{m}^{\mathcal{D}_{0}}}=0.38$. Therefore, the prior issue can be excluded. Since the same set of data shared by\\ FLRW case result in a reasonable conclusion, the lack of amount for OHD can also be neglected.

Finally, we believe that the prescription of $\kappa_{\mathcal{D}}$ should be modified, or some other scenarios should be considered. On the one hand, we can modify the prescription into the forms of Eqs. (\ref{eq:38}) and (\ref{eq:39}), which are not different from each other in this context but are distinct beyond the scaling solutions. On the other hand, we can add an appropriate positive constant to the expression of $\kappa_{\mathcal{D}}$ for complimenting the problem of not including positive possibility.

In order to further pinpoint the hinge of the issue, we explore the evolutions of $q^{\mathcal{D}}$ over effective scale factor $a_{\mathcal{D}}$ with best-fit values of both us and Larena et. al. It turns out that both results are similar in the tendency of the evolutions. The only differences lie in the different turning points for the slow evolutions and the present values of $q^{\mathcal{D}_0}$. The larger $n$ becomes, the earlier for the evolutionary curves change from fast to slow. In other words, despite of the constraints of the effective parameters, there are not much differences, which leaves both the constraints for mutual contradiction. It just proves our point that we must remain skeptical on the prescription of $\kappa_{\mathcal{D}}$ and consider other options as mentioned above.

\begin{acknowledgements}
We are grateful to Yu Liu for useful discussion. This work was supported by the National Science Foundation of China (Grants No. 11573006 ), National Key R\&D Program of China (2017YFA0402600)£¬the Fundamental Research Funds for the Central Universities£¬and the Special Program for Applied Research on Super Computation of the NSFC-Guangdong Joint Fund (the second phase).
\end{acknowledgements}

\bibliographystyle{spphys}
\bibliography{mybibfile}
\end{document}